\documentstyle[12pt,epsf]{article}
 \newcommand {\bi} {\bibitem}
 \newcommand {\be} {\begin{equation}}
\newcommand {\bea} {\begin{eqnarray} \nonumber }
\newcommand {\ee} {\end{equation}}
\newcommand {\eea} {\end{eqnarray}}
 \newcommand {\eps} {\epsilon}
 \newcommand {\si} {\sigma}
\newcommand {\de} {\delta}

 \newcommand {\al} {\alpha}

\newcommand {\lan} {\langle}
\newcommand {\ran} {\rangle}

\newcommand {\cC}  {{\cal C}}

\newcommand {\for} {\ \ \ \mbox{for}\ \ }

\def \form#1 {eq. (\ref{#1}) }
\def \parziale#1#2  {{\partial {#1} \over \partial {#2}}}

\begin{document}
\title{Testing replica predictions in  experiments}
\author{
Giorgio Parisi
\\ Dipartimento di Fisica, Universit\`a {\em La  Sapienza},\\ 
INFN Sezione di Roma I \\
Piazzale Aldo Moro, Rome 00185}
\maketitle
\begin{abstract}We review the predictions of the replica approach both for the statics and for the 
off-equilibrium dynamics.  We stress the importance of the Cugliandolo-Kurchan off-equilibrium 
fluctuation-dissipation relation in providing a bridge between the statics and the dynamics.  We 
present numerical evidence for the correctness of these relations.  This approach allows an 
experimental determination of the basic parameters of the replica theory.
\end{abstract}
\vskip.3cm

When we suddenly decrease the temperature in an Hamiltonian system, many interesting phenomena 
happen if the initial and the final temperatures correspond to different phases.  When the low 
temperature phase can be characterized by a simple order parameter (e.g.  the magnetization for 
ferromagnets) we find the familiar phenomenon of spinodal decomposition characterized by growing 
clusters of different phases.  There is a dynamical correlation length (i.e.  the size of the 
clusters), which increases as a power of the time $t$ after the quench, and the energy approaches 
equilibrium with power like corrections (e.g.  $E(t)\approx E_{\infty}+At^{-1/2}$).  In this region 
aging phenomena are also present \cite{B}.

The situation is more intriguing in the case of structural glasses and spin glasses where the low 
temperature phase cannot be characterized in term of a simple order parameter.  Remarkable 
progresses in understanding the off-equilibrium dynamics and its relations to the equilibrium 
properties has been done by noticing that a crucial off-equilibrium feature is the presence of 
deviations from the well known {\sl equilibrium} fluctuation-dissipation relations.  On the basis of 
analytic results for soluble models it has been conjectured that we can define a function $X(C)$, 
$C$ being an autocorrelation function at different times \cite{CUKU,FM,BCKM}.  This function 
characterizes the violations of the fluctuation-dissipation theorem (which is correct only at 
equilibrium).  It is remarkable that (at least in the case of spin glasses) the function $X(C)$ is 
equal to the function $x(q)$ ($q$ being the overlap of two spin configurations) which plays a 
central role in the equilibrium computation of the free energy \cite{mpv}.

Let us first recall the description of these glassy systems at equilibrium according to the 
predictions of the replica theory.  We consider a systems and we denote by $\cal C$ a generic 
configuration of the system.  For simplicity we will assume that there are no symmetry in the 
Hamiltonian, in presence of symmetries the arguments must be slightly modified.  It is useful to 
introduce an overlap $q(\cC,\cC')$.  There are many ways in which an overlap can be defined; for 
example in spin system we could define
\be
q={\sum_{i=1,N}\si_{i}\tau_{i}\over N},
\ee
$N$ being the total number of spins or particles and $\si$ and $\tau$ are the two spin 
configurations.  In a liquid a possibility is given by
\be
 q={\sum_{i=1,N}\sum_{k=1,N}f(x(i)-y(k))\over N},
\ee
where $f$ is a function which decays in a fast way at large distances and is substantially different 
from zero only at distances smaller that the interatomic distance ($x$ and $y$ are the two 
configurations of the system).

In the high temperature phase for very large values of $N$  the probability distribution 
of the overlap ($P_{N}(q)$) is given by
\be
P_{N}(q)\approx \delta (q-q^{*}).
\ee
In the low temperature phase $P_{N}(q)$ depends on $N$ (and on the quenched disorder, if it is 
present). When we average over $N$ we find a function $P(q)$ which is not a simple delta 
function.  In all known case one finds that
\be
P(q)=a_{m}\delta(q-q_{m})+a_{M}\delta(q-q_{M})+p(q),
\ee
where the function $p(q)$ does not contains delta 
function and its support is in the interval $[q_{m},q_{M}]$.

The non triviality of the function $P(q)$ (i.e.  the fact that $P(q)$ is not a single delta 
function and consequently $q$ is an intensive fluctuating quantity) is related to the existence of 
many  different equilibrium states.  Moreover the function $P_{N}(q)$ changes with $N$ and its 
statistical properties (i.e the probability of getting a given function $P_{N}(q)$) can be 
analytically computed \cite{mpv,BAPA}.

In this equilibrium description a crucial role is given by the function $x(q)$ defined
as
\be
x(q)=\int_{q_{m}}^{q}P(q')dq'\ .
\ee

In the simplest case the function $p(q)$ is equal to zero. i.e the function $P(q)$ has only two delta 
functions without the smooth part.  In this case, which correspond to one step replica symmetry 
breaking, there are many equilibrium states, labeled by $\al$, and the overlaps among two generic 
configurations of the same state and of two different states are respectively $q_{M}$ and $q_{m}$.  
The probability of finding a state with total free energy $f$ is proportional to
\be
\exp \left( m\beta(f- f_{R})\right), \label{REM}
\ee
where $f_{R}$ is a reference free energy and $m$ is the value of $x(q)$ in the interval 
$[q_{m},q_{M}]$.

In the more complicated situation where the function $p(q)$ is non zero, couples of different states 
may have different values of the overlaps.  The conjoint probability distribution of the states and 
of the overlaps can be described by formulae similar to eq.  (\ref{REM}), but more complex \cite{mpv}.

Although these predictions are quite clear, it is not so simple to test them for many reasons:
\begin{itemize}
\item They are valid  at thermal equilibrium,  a condition that is very difficult
to reach for this kind of systems.
\item Experimentally is extremely difficult to measure the values of the microscopic variables, i.e 
all the spins of the system at a given moment.  These measurements can be done only in numerical 
simulations, where the observation time cannot be very large.
\end{itemize}

A very important progress has been done when it was discovered \cite{CUKU} that the function $X$, 
which describes the violations of the fluctuation dissipation theorem, is equal to the function $x$ 
which is relevant for the statics.  This equality is very interesting because function $X(C)$ can be 
measured relatively easily in off-equilibrium simulations
\cite{FRARIE}.

The temperature dependence of the function $X(C)$ (or equivalently $x(q)$) is interesting 
also because rather different systems can be classified in the same universality class according to 
the behaviour of this function.  It has been conjectured long time ago that the equilibrium 
properties of glasses are in the same universality class of some simple generalized spin glass 
models
\cite{KWT,PARI}.

Let us be more precise.  We concentrate our attention on a quantity $A(t)$.  We suppose that the 
system starts at time $t=0$ from an initial condition and subsequently it remains at a fixed 
temperature $T$.  If the initial configuration is at equilibrium at a temperature $T'>T$, we observe 
an off-equilibrium behaviour.  We can define a correlation function
\be
C(t,t_{w}) \equiv \lan A(t_{w}) A(t+t_{w})\ran
\ee
and the response function
\be
G(t,t_{w}) \equiv \frac{ \de \lan A(t+t_{w})\ran}{\de \eps(t_{w})}{\Biggr |}_{\eps=0},
\ee
where we are considering the evolution in presence of a time dependent Hamiltonian in which we have
added the term
$ \int dt \eps(t) A(t) $.
 
The usual equilibrium fluctuation-dissipation theorem (FDT) tells us that
\be G^{eq}(t)= - \beta \frac{d C^{eq}(t)}{ dt}, \ee
where
\be
G^{eq}(t)=\lim_{t_w \to \infty} G(t,t_w), \ \ C^{eq}(t)=\lim_{t_w \to \infty} C(t,t_w).
\ee

It is convenient to define the integrated response:
\be
R(t,t_{w})=\int_{0}^{t} d\tau G(t-\tau,t_{w}+\tau),\ \ R^{eq}(t)=\lim_{t_w \to \infty} R(t,t_w),
\ee
$R(t,t_{w})$ is the response of the system at time $t+t_{w}$ to a field acting for a time $t$ 
starting at $t_{w}$.  The usual FDT relation becomes
\be
R^{eq}(t)= \beta (C^{eq}(t)-C^{eq}(0)).
\ee

\begin{figure}
\epsfysize=250pt
\epsffile{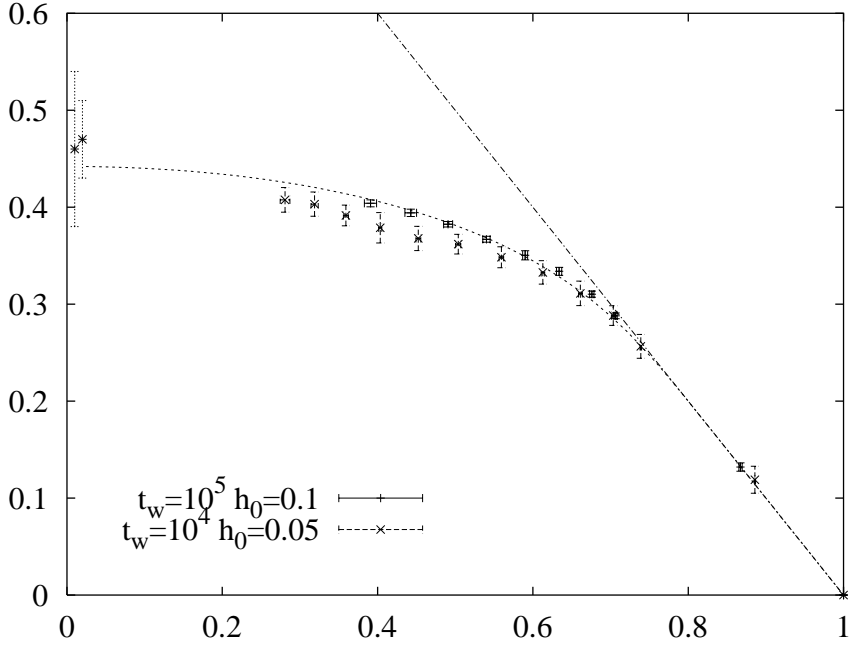}
\caption{The response $ R$ times $T$ versus $ C$ at $T=0.7$ for the three
dimensional Ising spin glass [10].  The curve is the prediction for function $R(C)$ obtained from the 
equilibrium data.  The straight line is the FDT prediction.  We have plotted the data of the two 
runs: $t_w=10^5$, and $t_w=10^4$.}
\protect\label{fig:gfdt}
\end{figure}

The off-equilibrium fluctuation-dissipation relation \cite{CUKU} states that the 
response function and the correlation function satisfy the following relation for large $t_w$:
\be
R(t,t_w)\approx \beta \int_{C(t,t_w)}^{C(0,t_{w})}X(C) dC.  \label{OFDR}
\ee
If we plot $R$ versus $\beta C$ for large $t_{w}$ the data collapse on the same 
universal curve and the slope of that curve is $-X(C)$.  The function $X(C)$ is system dependent and 
its form tells us  interesting information. in the case of three dimensional spin glasses.

We must distinguish two regions:
\begin{itemize}
\item A short time region where $X(C)=1$ (the so called FDT region) and $C$ belongs to the interval
$I$ 
(i.e. $C_1<C<C_2$.).

\item  A large time region (usually $t=O(t_w)$) where 
$C\notin I$ and $X(C)<1$.  In the same region the correlation function often satisfies an aging 
relation i.e $C(t,t_w)$ depends only on the ration $s \equiv t/t_{w}$ in the region where both $t$ 
and $t_{w}$ are large: $C(t,t_w)\approx C^{a}(t/t_{w})$.
\end{itemize}

In the simplest non trivial case, i.e.  one step replica symmetry breaking \cite{mpv,PARI} , the 
function $X(C)$ is piecewise constant, i.e.
\be
X(C)= m \for C \in I,\ \ X(C)= 1 \for C \notin I \label{ONESTEP}.
\ee
One step replica symmetry breaking for glasses has been conjectured in ref.  \cite{KWT}.

In all known cases in which one step replica symmetry holds, the quantity $m$ vanishes linearly with 
the temperature at small temperatures.  It often happens that $m=1$ at $T=T_{c}$ and $m(T)$ is 
roughly linear in the whole temperature range.  

Let us consider the case of spin glasses at zero magnetic field (in this case the replica symmetry 
is fully broken \cite{BOOK}).  The natural variable to consider is a single spin ($A=\si_{i})$).  
In this case the correlation $C(t,t_{w})$ is equal to the overlap among two configurations at time 
$t$ and $t_{w}$:
\be
C(t,t_{w})={\sum_{i=N}\si_{i}(t)\si_{i}(t_{w})\over N}.
\ee
The response function is just the magnetization in presence of an infinitesimal magnetic field.
In this case the situation is quite good because there are reliable simulations for the system at 
equilibrium \cite{BOOK}.

In fig.  (1) (taken from \cite {MPRR}) we plot the prediction for the function $R$ versus $C$, 
obtained at equilibrium (i.e.  using the equilibrium probability distribution of the overlaps, 
$P(q)$) by means of a simulation of a $16^3$ lattice using parallel 
tempering~\cite{HUKUNEMOTO,BOOK}.  The simulation has been done  with the help of the APE100 
supercomputer~\cite{APE} and involves the study of $900$ samples of a $ L=16$ lattice.

During the off-equilibrium simulations \cite {MPRR} in a first run without magnetic field the autocorrelation 
function has been computed.  In a second second run from $t=0$ until $t=t_w$ the magnetic field is 
zero and then (for $t \ge t_w$) there is an uniform magnetic field of small strength $h_0$.  The 
starting configurations were always chosen at random (i.e.  the system is suddenly quenched from 
$T=\infty$ to the simulation temperature $T$).

In fig.  (1) there are the results of the off-equilibrium simulations \cite {MPRR} where $t_w=10^5$ 
and $t_w=10^4$, with a maximum time of $5\cdot 10^6$ Monte Carlo sweeps.  The lattice size in was 
$64$, and $T=0.7$ (well inside the spin glass phase, the critical temperature is close to 1.0).  We 
plot the response function $ R$ times $T$ (in this case $R$ is equal to $m/h_0$) against $C(t,t_w)$.  
We have plotted also a straight line with slope $-1$ in order to control where the FDT is satisfied.  
Finally we have plotted two points, in the left of the figure, that are obtained with the infinite 
time extrapolation of the magnetization.

The agreement among the absolute theoretical predictions (no free parameters) coming from the 
statics and the dynamical numerical data is quite remarkable.  These data show the correctness of the 
identification of the functions $x$ of the statics and $X$ of the dynamics.

Let us now go to the case of glass forming materials.  I will present the data for  binary 
mixture of soft spheres \cite{HANSEN}.  Theoretically there have been many speculations the glass 
transition is described by one step replica symmetry breaking \cite{KWT,PARI,MEPA,FRAPA}.  Here the 
equilibrium properties are no so well known as in spin glasses, although there are some evidence 
that the homologous of the function $P(q)$ is non trivial \cite{BAPA}.  On the other side, as we 
shall see, off-equilibrium simulations \cite{PAGE} show that the function $X(C)$ seems to be given 
by the one step formula (\ref{ONESTEP}) with an approximate linear dependence of $m$ on the 
temperature.

\begin{figure}
\epsfxsize=250pt\epsffile{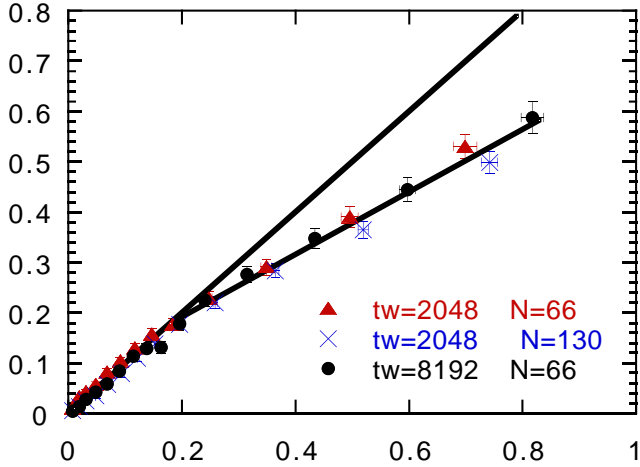}
\caption{ $R$ versus $\beta\Delta$ at $\Gamma=1.6$ for 
$t_{w}=8192$ and $t_{w}=2048$ at $N=66$ and for $t_{w}=2048$ at $N=130$.  The two straight lines 
have slope 1 and .62 respectively.}\label{UNO}
\end{figure}

We consider  a mixture of soft particles of different sizes.  Half of the particles 
are of type $A$, half of type $B$ and the interaction among the particles is given by the 
Hamiltonian:
\begin{equation}
H=\sum_{{i<k}} \left(\frac{(\si(i)+\si(k)}{|{\bf x}_{i}-{\bf x}_{k}|}\right)^{12},\label{HAMI}
\label{HAMILTONIAN}
\end{equation}
where the radius ($\si$) depends on the type of particles.  This model has been carefully studied in 
the past \cite{HANSEN,PAGE}.  The choice $\si_{B}/\si_{A}=1.2$ strongly inhibits crystallisation and 
the system goes into a glassy phase when it is cooled.  Using the same conventions of the previous 
investigators we consider particles of average radius $1$ at unit density.
It is usual to introduce the quantity $\Gamma 
\equiv \beta^{4}$.  For quenching from $T=\infty$ the glass transition is known to happen around
$\Gamma_c=1.45$
\cite{HANSEN}.

The best quantity we can measure to evidenziate off-equilibrium effects is the diffusion of the 
particles:
\be
\Delta(t,t_{w})\equiv {\sum_{i=1,N}\lan|{\bf x}_{i}(t_{w})-{\bf x}_{i}(t_{w}+t)|^{2} \ran \over N}.
\ee
The usual diffusion constant is given by $D=\lim_{t\to\infty}\Delta(t,t_{w})/t$.

The other quantity we measure is the response to a force.  At time $t_{w}$ we add to the Hamiltonian 
the term $\eps\ {\bf f} \cdot {\bf x}_{k}$, where $f$ is vector of squared length equal to $d=3$ and 
we measure the response
\be
R(t_{w},t)= {\partial \lan {\bf f} \cdot {\bf x}_{k}(t_{w}+t) \ran_{\eps}\over \partial \eps} {\Biggr 
|}_{\eps=0}
\approx { \lan {\bf f} \cdot {\bf x}_{k}(t_{w}+t)\ran_{\eps} \over \eps}
\ee
for sufficiently small $\eps$.
The usual fluctuation theorem tells that at equilibrium $\beta \Delta^{eq}(t)=R^{eq}(t)$. 

In the following we will look for the validity in the low temperature region of the generalized 
relation $\beta X(\Delta) =\partial R / \partial\Delta$.  This relation (with $X\ne 1$) can be valid 
only in the region where the diffusion constant $D$ is equal to zero.  Strictly speaking also in the 
glassy region $D\ne 0$, because diffusion may always happens by interchanging two nearby particles 
($D$ is different from zero also in a crystal); however if the times are not too large the value of 
$D$ is so small in the glassy phase that this process may be neglected in a first approximation.

\begin{figure}
\epsfxsize=250pt\epsffile{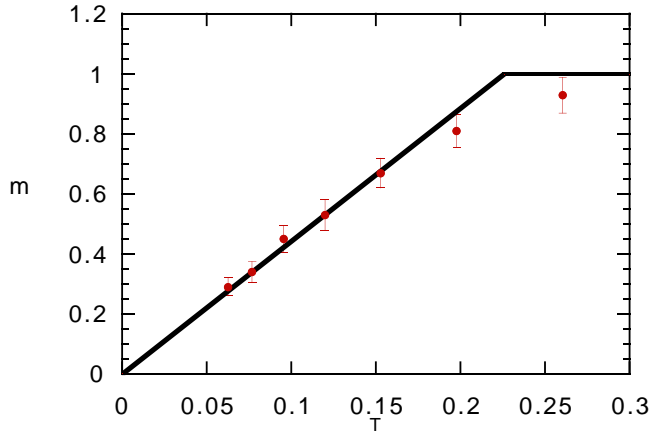}
\caption{ The quantity $m\equiv	{\partial R\over \partial \beta	\Delta}$ as $t_{w}=2048$ as
function of the temperature.  The straight line is the prediction of the approximation 
$m(T)=T/T_{c}$.}
\end{figure}\label{DUE}

The simulations we present are done using a Monte Carlo algorithm, which is a discretized form of a 
Langevin dynamics.  In fig.  \ref{UNO} we show $R$ versus $\beta\Delta$ at $t_{w}=2048$ and 
$t_{w}=8192$ for $\Gamma=1.6$ and $t\le 4t_{w}$ at $N=66$.  We also show the data for $t_{w}=2048$ 
at $N=130$.  We do not observe any significant systematic shift in this plot among three data sets.  
We distinguish two linear regions with different slope as expected from one step replica symmetry 
breaking.  The slope in the first region is compatible with 1, as expected from the FDT theorem, 
while the slope in the second region is near 0.62.  Also the data at different temperatures for all 
values of $\Gamma\ge 1.5$ show a similar behaviour.  The value of $R$, in the region where the FDT 
relation does not hold, can be very well fitted by a linear function of $\Delta$ as can be seen in 
fig.  \ref{UNO}.  The region where a linear fit (with $m<1$) is quite good corresponds to 
$t/t_w>0.2$.  

The fitted value of $m\equiv
\partial R/\partial (\beta
\Delta)$ is displayed in fig.  (3).  When $m$ becomes equal to 
1, the
fluctuation-dissipation theorem holds in the whole region and this is what happens at higher 
temperatures.  The straight line is the prediction of the approximation $m(T)=T/T_{c}$, using 
$\Gamma_{c}=1.45$.

All the results are in very good agreement with the theoretical expectations based on our knowledge 
extracted from the mean field theory for generalized spin glass models.  The approximation 
$m(T)=T/T_{c}$ seems to work with an embarrassing precision.  We can conclude that the ideas 
developed for generalized spin glasses have a much wider range of application than the models from 
which they have been extracted.  It likely that they reflect quite general properties of the phase 
space and therefore they can be applied in cases which are very different from the original ones.

The most interesting development would be to measure experimentally the function $X$ both in spin 
glasses and in structural glasses.  Clearly the most difficult task is the measurement of the 
fluctuations.  In spin glasses it is clear how it should be done: the measurement of the thermal 
fluctuations of the magnetization is a delicate, but feasible experiment.  In the case of structural 
glasses some ingenuity is needed in planning the experiments.  A open interesting possibility would 
to do the measurements in the case of rubber, where a transition with similar characteristics should 
take place.

\end{document}